\documentclass[aps,twocolumn,noshowpacs,superscriptaddress]{revtex4-1}
\usepackage{xspace}
\usepackage{bm}
\usepackage{diagbox}
\usepackage{graphicx}
\usepackage{xcolor,url}
\usepackage{amsmath,amssymb,amsfonts,amsthm,physics}
\usepackage[colorlinks=true,linkcolor=blue,anchorcolor=red,citecolor=blue,urlcolor=blue]{hyperref}
\usepackage{multirow,makecell}

\makeatletter
\AtBeginDocument{\let\LS@rot\@undefined}
\makeatother

\def \K {\hat{\mathcal{K}}}
\def \Z {\mathbb{Z}}
\def \H {\mathcal{H}}
\def \k {\bm{k}}
\def \kk {\mathsf{k}}

\def \TT {\hat{\mathcal{T}}}
\def\T{\mathcal{T}}
\def \CC {\hat{\mathcal{C}}}

\def \II {\hat{\mathcal{I}}}

\begin{document}

\title{Tensor Theory for Higher Dimensional Chern Insulators with Large Chern Numbers}

\author{Kai Wang}
\email[These authors contributed  equally  to this work.]{}
\affiliation{National Laboratory of Solid State Microstructures and Department of Physics, Nanjing University, Nanjing 210093, China}

\author{Jia-Xiao Dai}
\email[These authors contributed  equally  to this work.]{}
\affiliation{National Laboratory of Solid State Microstructures and Department of Physics, Nanjing University, Nanjing 210093, China}

\author{L. B. Shao}
\email[]{lbshao@nju.edu.cn}
\affiliation{National Laboratory of Solid State Microstructures and Department of Physics, Nanjing University, Nanjing 210093, China}
\affiliation{Collaborative Innovation Center of Advanced Microstructures, Nanjing University, Nanjing 210093, China}

\author{Shengyuan A. Yang}
\affiliation{Research Laboratory for Quantum Materials, Singapore University of Technology and Design, Singapore 487372, Singapore}

\author{Y. X. Zhao}
\email[]{zhaoyx@nju.edu.cn}
\affiliation{National Laboratory of Solid State Microstructures and Department of Physics, Nanjing University, Nanjing 210093, China}
\affiliation{Collaborative Innovation Center of Advanced Microstructures, Nanjing University, Nanjing 210093, China}

\begin{abstract}
	Recent advances in topological artificial systems open the door to realizing topological states in dimensions higher than the usual three-dimensional space.
Here, we present a ``tensor product" theory, which offers a method to construct Chern insulators with arbitrarily high dimensions and Chern numbers. Particularly, we show that the tensor product of 
a $d_A$D Chern insulator $\langle \H_A^{(\kappa_A)}, C_A\rangle$ with a $d_B$D Chern insulator $\langle \H_B^{(\kappa_B)}, C_B\rangle$ leads to a $(d_A+d_B)$D Chern insulator $\langle \H_{A B}^{(\kappa_A\star \kappa_B)},-2C_AC_B\rangle $, where in the brackets, $\H^{(\kappa)}$ is the $d$D Hamiltonian with $d$ even, $C$ is the
corresponding $(d/2)$th Chern number, and $\kappa$ labels  the five non-chiral Altland-Zirnbauer symmetry classes A, AI, D, AII and C. The four real classes AI, D, AII and C form a Klein four-group under the multiplication ``$\star$'' with class AI the identity, and class A is the zero element. Our theory leads to novel higher-dimensional topological physics. (i) The construction can generate large higher-order Chern numbers, e.g., for some cases the resultant classification is $8\mathbb{Z}$. (ii) Fascinatingly, the boundary states feature flat nodal hypersurfaces with nontrivial Chern charges.  For the constructed $(d_A+d_B)$D Chern insulator, a boundary perpendicular to a direction of $\H_A$ generically hosts $|C_A|$ $d_B$D nodal hypersurfaces, each of which has topological charge $\pm 2C_B$. Under perturbations, each nodal hypersurface bursts into stable unit nodal points, with the total Chern charge conserved. Examples are given to demonstrate our theory, which can be experimentally realized in artificial systems such as acoustic crystals, electric circuit arrays, ultracold atoms, or mechanical networks. 
\end{abstract}
	
\maketitle
	
\section{Introduction}
Although our daily experience is limited to three spatial dimensions, it does not prevent us from exploring interesting physics  of higher dimensional spaces~\cite{Zhang_2001}. In the research on topological matters, a milestone is the classification table for topological insulators/superconductors in the eight real Altland-Zirnbauer (AZ) symmetry classes, which was found to exhibit an eightfold periodicity in dimensions \cite{AZ-Classification,Schnyder08prb,Kitaev2009AIP}. Later, a similar eightfold periodicity was also observed for the classifications of (semi)metals and nodal superconductors \cite{ZhaoYX2013prl,ZhaoWang16Aprprl,ShinseiRyu-RMP}. 

Remarkably, the recent progress in artificial systems has pushed the higher-dimensional matters beyond purely theoretical interest~\cite{Kraus2013prl,Price2015prl,Lohse2018,Zilberberg2018,ZhangShuang2019science,ZhangShuang2019prl,Ezawa2019prb,WangDY2019np,Yu_Zhao_NSR,WangYou_2020,ZhangWX2020prb,ZhangShuang2021nature}. Compared with condensed matter systems, artificial crystals,
such as photonic/ascoustic crystals, cold atoms in optical lattices, electric-circuit arrays, and mechanical networks, have the advantage of 
flexibly engineering tight-binding models beyond the three-dimensional (3D) limit. 
For instance, the four-dimensional (4D) quantum Hall effect has been successfully simulated by both ultracold atoms and photonic crystals \cite{Zilberberg2018,Lohse2018}. A $4$D Chern-insulator model in class AI with the $2$nd Chern number of $2$ has been projected onto a two-dimensional (2D) circuit { board} and simulated by a periodic LC network \cite{Yu_Zhao_NSR}. These achievements have stimulated tremendous interest on higher-dimensional topological matters and call for systematic investigation of their properties.

In this paper, motivated by the experimental advances, we {develop} a method, termed ``tensor product", to systematically construct higher-dimensional Chern insulators (ChIs) from lower-dimensional ones. In the AZ symmetry classes, ChIs can occur in five non-chiral classes, namely 
class A, AI, D, AII and C \cite{AZ-Classification,Schnyder08prb,Kitaev2009AIP,ZhaoYX2013prl}.
Our tensor product induces a multiplication ``$\star$" for these five classes, under which the four real classes AI, D, AII and C form a Klein four-group with AI the identity, and class A is the zero element, as illustrated in Table~\ref{symmetry-class}. For a $d$D ChI $\H$ in class $\kappa$, it has a nontrivial $p$th Chern number $C^p(\H)$ (with $p=d/2$), which is an integer in $\Z$ or an even integer in $2\Z$, depending on $d$ and $\kappa$. We show that the tensor product of two ChIs $\H_A$ and $\H_B$  gives a $(d_A+d_B)$D Chern insulator in class $\kappa_A\star\kappa_B$, whose $(p_A+p_B)$th Chern number is equal to $-2C_AC_B$. Particularly, the tensor product of two ChIs with $2\Z$ classification leads to $8\Z$, i.e., the resulting Chern number is a multiple of 8. Thus, our approach can naturally generate ChIs with large Chern numbers. For instance, we explicitly show the tensor product of two $2$D ChIs in class C gives a $4$D ChI in class AI with second Chern number $8$.

Recently, topological nodal surfaces have attracted both theoretical and experimental interests \cite{Zhong2016,Liang2016,Tomas2017prb,xiao2017topologically,ZhaoYang2018prb,Yu2019b,YangYH2019nc,XiaoMeng2020sciadv,Schnyder2021nature}. Fascinatingly, the constructed ChIs by the tensor product automatically have nodal-hypersurface boundary states. For the constructed $(d_A+d_B)$D ChI, a boundary perpendicular to a direction of $\H_A$ ($\H_B$) generically hosts $|C_A|$ $d_B$D ($|C_B|$ $d_A$D) nodal hypersurfaces, each of which has topological charge $\pm 2C_B$ ($\pm 2C_A$). The total topological charge of these nodal hypersurfaces is therefore consistent with the bulk Chern number $-2C_AC_B$. As we shall see, considering the boundary nodal hypersurfaces naturally explains the factor $2$ appearing in $-2C_AC_B$. These nodal hypersurfaces are not topologically stable. Under generic perturbations, each nodal hypersurface bursts into stable nodal points with unit Chern charges $\pm 1$, with the total Chern charge conserved. Concrete models are constructed to demonstrate our theory, which can be readily realized using artificial systems.

\section{Tensor product of ChIs}
Let us consider two ChIs with momentum-space Hamiltonians $ \H_A(\k^A) $ and $ \H_B(\k^B) $, respectively, which may have different spatial dimensions $2p$ and $2q$ and therefore different orders of Chern numbers $C^{p}$ and $C^{q}$, respectively. Their ``tensor products" are constructed as
\begin{equation}\label{tensoring}
	\H_{AB}(\k)= \H_A(\k^A)\otimes\H_B(\k^B),
\end{equation}
where $ \k=(\k^A,\k^B) $, and $\otimes$ is the ordinary tensor product of two matrices. Thus, $\H_{AB}(\k)$ describes a $(2p+2q)$D system with first $2p$ dimensions coming from $\H_A$ and the other $2q$ dimensions from $\H_B$. This construction is inspired by the tensor product of vector bundles in mathematics \cite{Karoubi,Atiyahbook}. Nevertheless, in mathematics, there is no notion of a conduction/valence state, but in physics, we have to focus on a spectral gap (assumed to be centered at zero energy) and make a distinction between conduction- and valence-state bundles. Then, the operation in (\ref{tensoring}) would mix the valence-state bundles of $A$ with
 the conduction-state bundles of $B$ and vice versa. This is a crucial point in our following discussion.

In the framework of tenfold classification, as we have mentioned, a ChI belongs to one of the five non-chiral AZ classes, A, AI, D, AII and C, because chiral symmetry would constrain the Chern number to be zero. {The proof can be found in Appendix \ref{AppendixE}.} The tensor product \eqref{tensoring} induces a multiplication $\star$ on these five classes. Labeling the class of $\H_\alpha$ by $\kappa_\alpha$, $\alpha\in\{A,B\}$, then $\H_{AB}$ is in the class $\kappa_A\star\kappa_B$. For any real class $\kappa$, namely $\kappa\ne\ $A, it has an anti-unitary symmetry $\mathcal{A}_\kappa=U_\kappa\II\K$, where $U_{\kappa}$ is a unitary operator, $\K$ the complex conjugation and $\II$ the inversion of $\k$. The symmetry constrains the Hamiltonian through $U_\kappa\H^*(\k)U^\dagger_{\kappa}=\epsilon_\kappa \H(-\k)$, and $U_\kappa U^*_\kappa=\eta_{\kappa}I$, where $\epsilon_\kappa=\pm 1$, $\eta_\kappa=\pm 1$ and $I$ is the identity operator. The four real classes can be identified by the values of $(\epsilon_\kappa, \eta_{\kappa})$, i.e.,
$\mathrm{AI}=(1,1),\mathrm{D}=(-1,1),\mathrm{AII}=(1,-1),\mathrm{C}=(-1,-1)$.
Under the tensor product \eqref{tensoring}, the symmetry operator for $\H_{AB}$ is $\mathcal{A}_{\kappa_A\star \kappa_B}=U_{\kappa_A}\otimes U_{\kappa_B}\II\K$, where $\II$ inverses both $\k^A$ and $\k^B$ and $\K$ operates on the whole $\H_{AB}$. Accordingly, we find $\epsilon_{\kappa_A\star \kappa_B}=\epsilon_{\kappa_A}\epsilon_{\kappa_B}$ and $\eta_{\kappa_A\star \kappa_B}=\eta_{\kappa_A}\eta_{\kappa_B}$. Thus, the four real classes under the multiplication are equivalent to the Klein four-group with the multiplication table explicitly given in Table~\ref{symmetry-class} \cite{Kleingroup}. Moreover, class A can be regarded as the zero element, since the tensor product with $\H_A$ in class A can break the symmetry of $\H_B$, as indicated in Table~\ref{symmetry-class}.

\begin{table}[t]
	\begin{tabular}{c|ccccc}
		\hline\hline
		& A &AI&D&AII&C\\
		\hline
		A&A&A&A&A&A\\
		AI& A&AI&D&AII&C \\
		D&A&D&AI&C&AII\\
		AII&A&AII&C&AI&D\\
		C&A&C&AII&D&AI\\
		\hline\hline
	\end{tabular}
	\caption{The multiplication table of non-chiral AZ symmetry classes. Class A is the zero element, and real classes AI, D, AII and C form the Klein four-group $\Z_2\times\Z_2$ with AI the identity element. \label{symmetry-class}}
\end{table}


\section{The Chern-number formula}

We proceed to derive the $(p+q)$th Chern number $C^{p+q}_{AB}$ of $\H_{AB}$ resulting from our construction in Eq.~\eqref{tensoring}. The result is an elegant formula:
\begin{equation}\label{Main-Result}
	C^{p+q}_{AB}=-2C^{p}_AC^{q}_B,
\end{equation}
where $ C^{p}_A $ and $C^{q}_B $ are the $ p $th and $ q $th Chern numbers of $ \H_{A} $ and $ \H_{B} $, respectively.

To prove \eqref{Main-Result}, we need to first look into the spectrum of $\H_{AB}$. Let $|i_{+},\k^\alpha\rangle$ ($|i_{-},\k^\alpha\rangle$) be the $i$th conduction (valence) eigenstate of $\H_\alpha(\k^\alpha)$, i.e.,
$\H_{\alpha}(\k^{\alpha})\ket{i_{\pm},\k^{\alpha}}=E_{i_{\pm},\k^{\alpha}}\ket{i_{\pm},\k^{\alpha}}$, with $E_{i_{+},\k^{\alpha}}>0$ and $E_{i_{-},\k^{\alpha}}<0$. Suppose there are $ N^+_{\alpha} $ conduction bands and $ N^-_{\alpha} $ valence bands for $\H_\alpha(\k^{\alpha}) $. Then, the spectrum of $ \H_{AB}(\k) $ is given by
\begin{equation}\label{Sch-Eq}
	\H_{AB}(\k)\ket{i_s,j_{s'},\k}=E_{i_s,\k^A}E_{j_{s'},\k^B}\ket{i_s,j_{s'},\k},
\end{equation}
where $\ket{i_s,j_{s'},\k}:=\ket{i_s,\k^A}\otimes\ket{j_{s'},\k^B} $, and $s,s'=\pm$. Hence, there are totally $(N^{+}_AN^{-}_B+N^{-}_AN^{+}_B)$ valence bands for $ \H_{AB}(\k) $, and the valence states are given by
\begin{equation}\label{AB_valence}
\ket{i_{-},\k^A}\otimes\ket{j_{+},\k^B},\ \ket{i_{+},\k^A}\otimes\ket{j_{-},\k^B}.
\end{equation}



Then, we can directly compute the Berry connection  for these valence bands. For simple notations, let us denote $ \k^A=(k_1,k_2,\cdots,k_{d_A}) $ and $ \k^B=(k_{\bar{1}},k_{\bar{2}},\cdots,k_{\bar{d}_B}) $. The Berry connection $A_{a}(\k)$ with $a=\mu,\bar{\nu}$ can be written as
\begin{equation}\label{Berry-Connection}
	\begin{split}
		A_{\mu}(\k)&=(a_{\mu}^{A,-}\otimes I^{B,+})\oplus (a_{\mu}^{A,+}\otimes I^{B,-}),\\
		A_{\bar{\nu}}(\k)&=(I^{A,-}\otimes a_{\bar{\nu}}^{B,+})\oplus (I^{A,+}\otimes a_{\bar{\nu}}^{B,-}).
	\end{split}
\end{equation}
Here, $ a^{\alpha,\pm} $ are the Berry connections for the conduction and valence bands of system $ \alpha $, and $ I^{\alpha,\pm} $ are the identity matrices. The Berry curvature $F_{ab}$ obtained from the Berry connection $A_{a}$ in \eqref{Berry-Connection} is
\begin{equation}\label{Berry-Curvature}
	\begin{split}
		F_{\mu\nu}&=(f_{\mu\nu}^{A,-}\otimes I^{B,+})\oplus (f_{\mu\nu}^{A,+}\otimes I^{B,-}),\\
		F_{\bar{\mu}\bar{\nu}}&=(I^{A,-}\otimes f_{\bar{\mu}\bar{\nu}}^{B,+})\oplus (I^{A,+}\otimes f_{\bar{\mu}\bar{\nu}}^{B,-}),\\
		F_{\mu\bar{\nu}}&=F_{\bar{\mu}\nu}=0.
	\end{split}
\end{equation}
Here, $ f^{\alpha,\pm} $ are the Berry curvatures of the conduction and valence bands for system $ \alpha $, respectively.
It is convenient to introduce  $ F_A=F_{\mu\nu}dk_{\mu}\wedge dk_{\nu}/2 $ and $ F_B=F_{\bar{\mu}\bar{\nu}}dk_{\bar{\mu}}\wedge dk_{\bar{\nu}}/2 $, and accordingly the total Berry curvature $F=F_A+F_B$. Recalling the definition of $n$th Chern number \cite{NAKAHARA},
\begin{equation}\label{Chern-number}
	C_{n}[F]=\frac{1}{n!}\left(\frac{i}{2\pi}\right)^{n}\int_\text{BZ}\tr F^{n},
\end{equation}
where the integral is over the whole Brillouin zone (BZ).
Then, using the fact that $[F_A,F_B]=0$, we find
\begin{equation}\label{CH}
	C_{AB}^{p+q}=\frac{1}{p!q!}\left(\frac{i}{2\pi}\right)^{p+q}\int_{\text{BZ}}\tr [(F_A)^{p}(F_B)^{q}].
\end{equation}
Further substituting \eqref{Berry-Curvature} into \eqref{CH} , we arrive at
\begin{equation}
	C_{AB}^{p+q}=C_{A,-}^pC_{B,+}^q+C_{A,+}^pC_{B,-}^q.
\end{equation}
Here, the subscript ``$\pm$" was added to distinguish the Chern numbers of conduction and valence bands. Derivation details can be found in Appendix \ref{AppexdixA}. Since $C:=C_{-}=-C_{+}$, the formula \eqref{Main-Result} is proved. In retrospect, we see that the two sets of valence bands in \eqref{AB_valence} equally contribute to $C^{p+q}_{AB}$, which explains the factor $2$ in \eqref{Main-Result}.

\begin{table}
	\centering
	\begin{tabular} {cc|cc|cc}
		\hline\hline
		\multicolumn{2}{c|}{\multirow{2}{*}{\diagbox{$~~~~A$}{ $B~~~~$}}} & \multicolumn{2}{c|}{2D} & \multicolumn{2}{c}{4D}\\ 
		\multicolumn{2}{c|}{} & D($ \Z $) & C($ 2\Z $) & AI($ 2\Z $) & AII($ \Z $) \\ \hline
		\multirow{2}{*}{2D} & D($ \Z $) & AI($ 2\Z $) & AII($ 4\Z $) & D($ 4\Z $) & C($ 2\Z $)  \\ & C($ 2\Z $) & AII($ 4\Z $) & AI($ 8\Z $) & C($ 8\Z $) & D($ 4\Z $) \\ \hline
		\multirow{2}{*}{4D} & AI($ 2\Z $) & D($ 4\Z $) & C($ 8\Z $) & AI($ 8\Z $) & AII($ 4\Z $)  \\ & AII($ \Z $) & C($ 2\Z $) & D($ 4\Z $) & AII($ 4\Z $) & AI($ 2\Z $) \\ \hline\hline
	\end{tabular}
	\normalsize
	\caption{The classification results of tensor products of Hamiltonians in four real non-chiral AZ symmetry classes. $\kappa(n\Z)$ denotes class $\kappa$ with classification $n\Z$. The dimension for each block entry is the sum of the row and column dimensions. }\label{Topological-Invariant}
\end{table}

Applying the multiplication Table~\ref{symmetry-class} of symmetry classes and the formula \eqref{Main-Result}, we can obtain the classification for the tensor products of any two ChIs. For the products among
$2$D and $4$D ChIs, the results are explicitly shown in Table~\ref{Topological-Invariant}. We observe that some entries are $8\Z$, i.e., the most elementary tensor product has Chern number $8$.

Although only the product of two systems is considered above, the tensor product can be repeatedly applied to give the product of any number of ChIs. Suppose there are $ N $ ChIs with Hamiltonians $ \H_1,\H_2,\cdots,\H_N $ of $ 2p_1,2p_2,\cdots,2p_N $ dimensions, respectively. Their tensor product is given by $\H(\k)= \H_1(\k^1)\otimes\H_2(\k^2)\otimes\cdots\otimes\H_N(\k^N)$, with $\k=(\k^1,\k^2,\cdots,\k^N)$. The resulting $\H$ has the $(p_1+\cdots+p_N)$th Chern number given by
\begin{equation}
	C^{p_1+\cdots+p_N}=(-2)^{N-1}\prod_{i=1}^{N} C^{p_i},
\end{equation}
which can be derived from repeatedly applying \eqref{Main-Result}.
Here, $ C^{p_i} $ is the $ p_i $th Chern number for $ \H_i $.

\section{Bulk-boundary correspondence}

The tensor-product ChI \eqref{tensoring} should have topological nontrivial boundary states according to the bulk-boundary correspondence ~\cite{Hatsugai1993prl,ZhaoYXWang14Septprb,Prodan-TI}. 


For example, let us first consider that $ \H_{A}(\k^A) $ has two dimensions, and we focus on the boundary of $\H_{AB}$ normal to the first dimension of $ \H_{A}(\k^A) $. Then the boundary BZ is parameterized by $(k_y,\k^B)$. Generically, there are $|C_A|$ chiral edge states for $H_A(k_y)$, where $H_A(k_y)$ is the Hamiltonian of system $A$ under the open boundary condition. These edge bands cross zero energy at $|C_A|$ points $k_y^a$ with $a=1,\cdots, |C_A|$. Around each $k_y^a$, the edge effective theory is just $h_a(q_a)=v_a q_y^a$ with $q^a_y$ the momentum deviation from $k_y^a$ and $v_a$ is the velocity of the mode.  For the tensor-product system $H_{AB}(k_y,\k^B)$, each $k_y^a$ is associated with a $(2q)$D nodal hypersurface, since $H_{AB}(k_y^a,\k^B)=0$ for each $k_y^a$. The boundary effective theory is given by $v_a q_y^a \H_B(\k^B)$. Hence, away from $k_y^a$, the boundary effective theory is gapped. Then, we can choose two hyperplanes at $q_{+}^a>0$ and $q_{-}^a<0$ to enclose the nodal hypersurface, and the topological charge of the nodal hypersurface is given by $C_a=C_{-,q_{+}}-C_{-,q_{-}}$. Here, $C_{-,q_+}=C_B$, but $C_{-,q_{-}}=-C_B$, since the valence states of $v_a q_{-}\H_B(\k^B)$ are the conduction states of $\H_B(\k^B)$. Thus, the nodal hypersurface must have a charge of $\pm 2C_B$.

The argument can be readily generalized to higher dimensional $\H_A(\k^A)$, where the boundary effective theory is of the form $v_i\bm{q}_i^a\gamma^i\otimes \H_B(\k^B) $ for each boundary nodal point $a$ of $H_A(\k^A_\perp)$. {Note that $\gamma^i$'s are $2^{p-1}$-dimensional Dirac matrices, satisfying Clifford algebra $\{\gamma^i,\gamma^j\}=\delta_{ij}$.} Then, we can enclose each nodal hypersurface by $S^{2p-2}\times T^{2q}$, where $S^{2p-2}$ encloses the nodal point of $v_i\bm{q}_i^a\gamma^i$ and $T^{2q}$ is the Brillouin torus of $\H_{B}$. The calculation of topological charge over $S^{2p-2}\times T^{2q}$ is exactly as what we did for the tensor product of ChIs, and therefore is equal to $\pm 2C_B$. Since there are $|C_A|$ such hypersurfaces, the
total topological charge of these nodal hypersurfaces is consistent with the bulk Chern number \eqref{Main-Result}. Here, the factor $2$ in the topological charges of boundary nodal hypersurfaces also provides a physical explanation for the factor $2$ in the bulk Chern number~\eqref{Main-Result}.

It is worth noting that these nodal hypersurfaces are not stable. In general, with symmetry-preserving perturbations, each of them can burst into nodal points with unit ($p+q-1$)th Chern numbers. Nevertheless, the total topological charge is conserved under the process.

\section{Two examples}

We demonstrate our theory by two examples.

\begin{figure}[t]
	\includegraphics[scale=0.38]{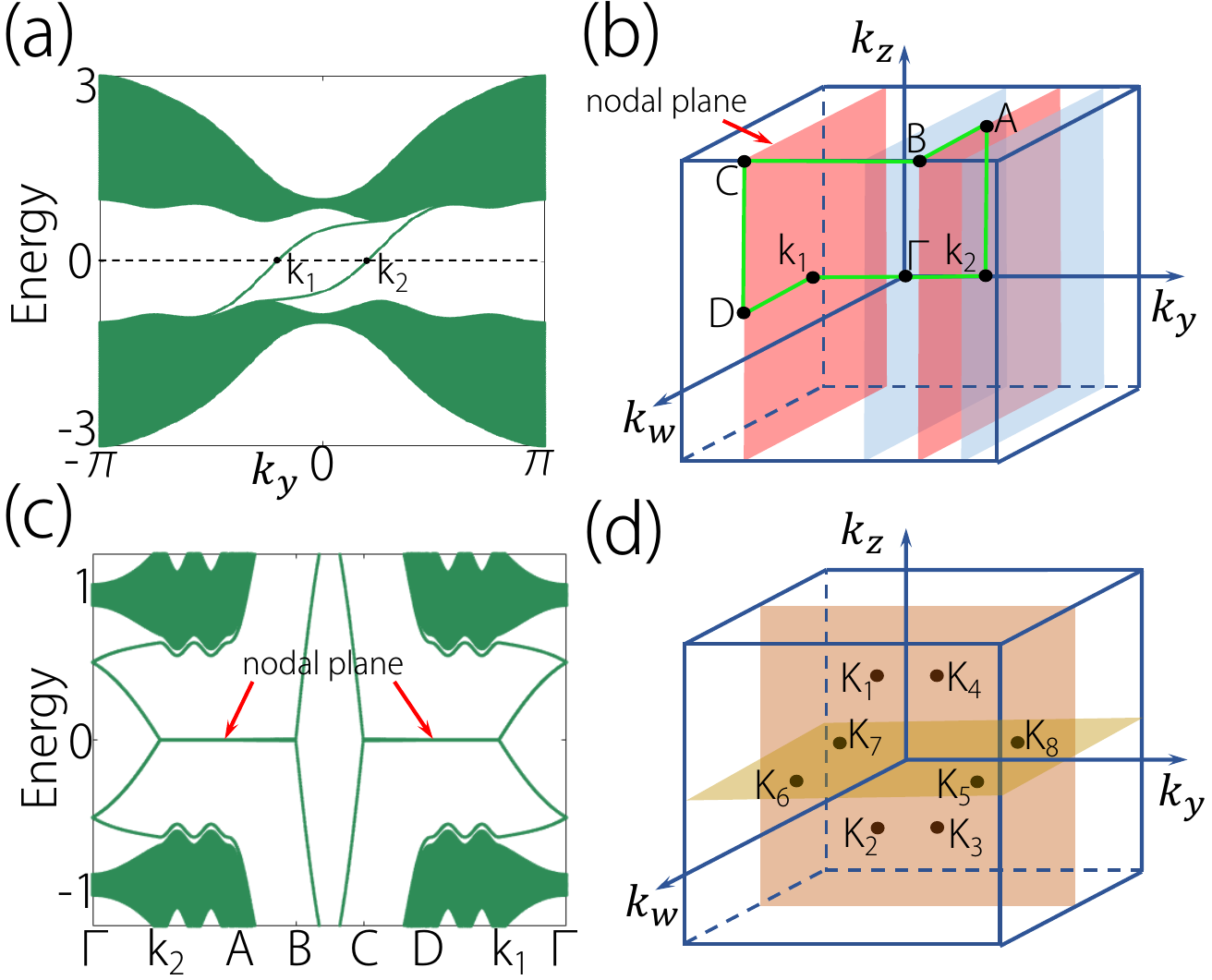}
	\caption{(a) The energy spectrum of $ \H_{\text{2D}}^{\text{C}} $, Eq.~\eqref{H-C}, with $ m=1 $. The edge is opened normal to the $x$ direction.  There are two chiral modes at each edge. (b) The two nodal planes marked in pink in the 3D boundary BZ for the Hamiltonian \eqref{H-AI-4D}. Each nodal plane is enclosed by two gapped planes marked in blue. The boundary is normal to the $ x $ direction. (c) The energy spectrum over the green line of (b). (d) The eight Weyl points $ K_1, K_2,\cdots, K_8 $ in the boundary BZ. Under the perturbation $ \Delta\H_2 $ with $ \Delta=0.2 $, each nodal plane bursts into four right-handed Weyl points. \label{fig1} }
\end{figure}

\begin{figure}[t]
	\includegraphics[scale=0.38]{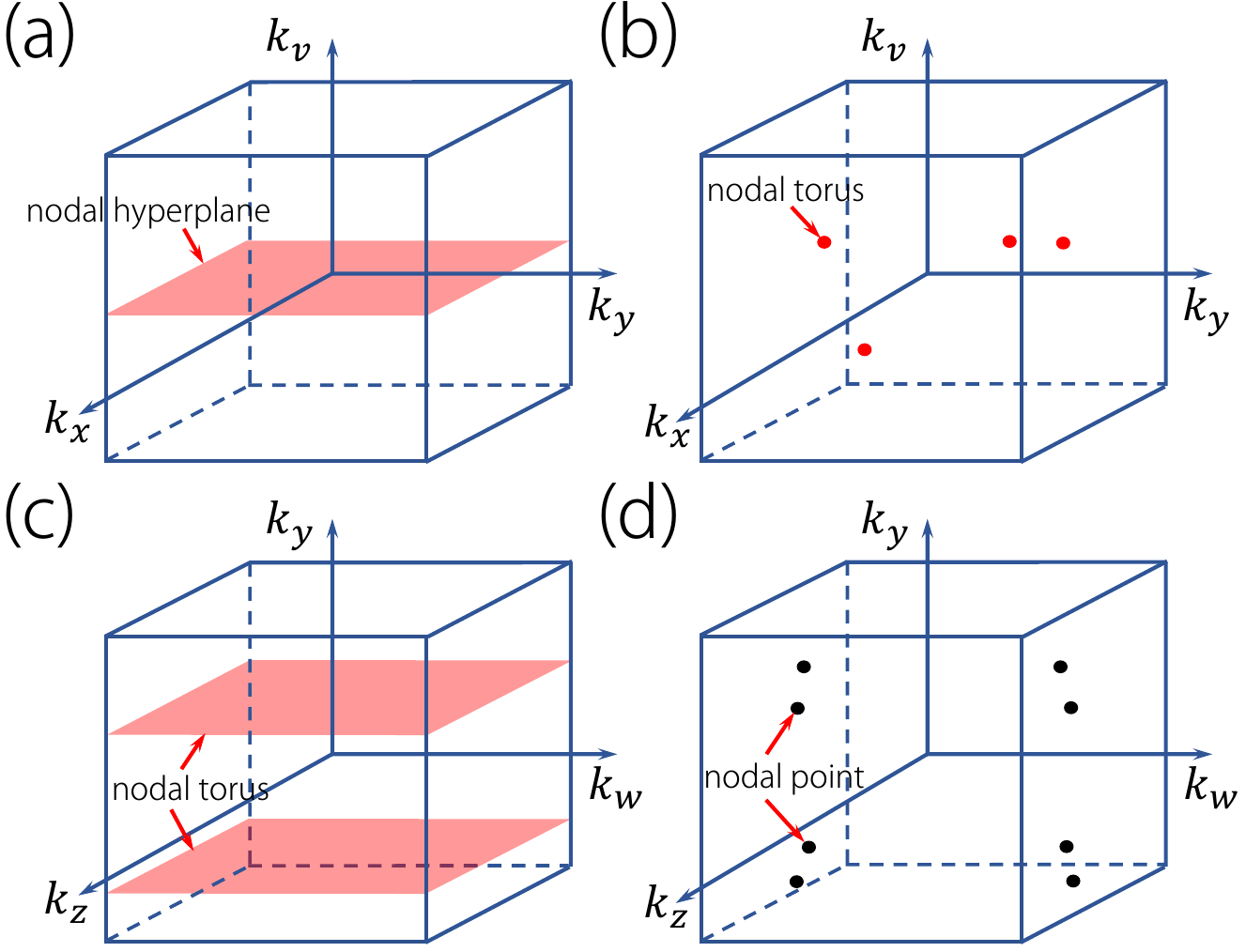}
	\caption{(a) The nodal hypersurface in the boundary sub-BZ coordinated by $k_x$, $k_y$, and $k_v$. (b) The nodal hypersurface bursts into four $2$D nodal torus under the perturbation $\alpha \sigma_1\otimes \H_{2\mathrm{D}}^\mathrm{C}(k_z,k_w)\otimes\sigma_2$. (c) Two nodal tori in the boundary sub-BZ coordinated by $k_z$, $k_w$ and $k_y$. (d) Each nodal torus bursts into four nodal points under the perturbation $\beta\H^\mathrm{C}_{\mathrm{2D}}(k_x,k_y)\otimes\sigma_1\otimes\sigma_2 +\delta \sigma_1\otimes\sigma_1\otimes\sigma_2$. Each nodal point has the second Chern charge $1$. {The parameters are taken as $\alpha=\beta=\delta=0.2$} \label{fig2}}
\end{figure}

We first construct the product of two $2$D class C ChIs with the first Chern number 2, which gives a $4$D ChI in class AI with the second Chern number $-8$.
The AZ symmetry class C  has $2\Z$ topological classification in two dimensions~\cite{Schnyder08prb}. We consider such a model,
\begin{equation}\label{H-C}
	\begin{split}
		\H_{\text{2D}}^{\text{C}}(\k)=&(\sin^2 k_x-\sin^2 k_y)\sigma_1+\sin k_x\sin k_y\sigma_2
		\\
		&+(m-\cos k_x-\cos k_y)\sigma_3
	\end{split}
\end{equation}
with the particle-hole symmetry $ \CC=\sigma_2\II\K $, where $\sigma$'s are the Pauli matrices. When $0<m<2$ ($-2<m<0$), the first Chern number $ C^1=-2 $ ($C^1=2$). For $m=1$, there are two chiral edge modes with two Fermi points at $k_y=\kk_{1,2} $, as shown in Fig.~\ref{fig2}(a). The tensor product of two copies of the Hamiltonian \eqref{H-C} is given by
\begin{equation}\label{H-AI-4D}
	\H_{\text{4D}}^{\text{AI}}(\k)=\H_{\text{2D}}^{\text{C}}(k_x,k_y)\otimes \H_{\text{2D}}^{\text{C}}(k_z,k_w),
\end{equation}
which belongs to class AI with the time-reversal operator $ \TT=(\sigma_2\otimes\sigma_2)\II\K $. With $m=1$ for both copies, we have the second Chern number $C^2_{AB}=-8$ for $\H_{\text{4D}}^{\text{AI}}$, according to our formula \eqref{Main-Result}. This value is also confirmed by directly calculating the second Chern number for $\H_{\text{4D}}^{\text{AI}}$, which can be found in Appendix \ref{AppexdixB}.

For a boundary normal to the $x$ direction, the $3$D boundary BZ is coordinated by $k_y$, $k_z$ and $k_w$. As indicated in Fig.~\ref{fig1}(b), there are two nodal planes (tori) at $ k_y=\kk_{1} $ and $\kk_2$, respectively. Fig.~\ref{fig1}(c) shows the energy spectrum over the green path in Fig.~\ref{fig1}(b), where the flat spectrum arises because its momentum lies on the nodal plane.  For each nodal plane, the topological charge is just the difference of first Chern numbers $C_{\pm}^1$ of two planes on the two sides of the nodal plane [see Fig.~\ref{fig1}(b)]. Direct calculation shows that $C_{\pm}^1=\mp 2$, and therefore the topological charge $C^1_{\kk_i}=-4$ for $i=1,2$.

The topological charges of the two nodal planes can be further verified by adding perturbations to break them into Weyl points. To see this, we add the term $\Delta\H_2=\Delta \sigma_1\otimes\sigma_1 $ to the tensor-product Hamiltonian \eqref{H-AI-4D}, which preserves  $\T$. Then , each nodal plane would burst into four right-handed Weyl points, conserving the total topological charge [see Fig.~\ref{fig1}(d)]. The detailed calculation can be found in Appendix \ref{AppexdixC}.

In the second example, we perform a tensor product of the $4$D ChI in \eqref{H-AI-4D} (class AI)  with a $2$D ChI in class D~\cite{XLQ08prb}. The resultant $6$D ChI is in class D according to Table~\ref{symmetry-class}, and the Hamiltonian is given by
\begin{equation}
	\H_{6\mathrm{D}}^{\mathrm{D}}(\k)=\H^{\mathrm{AI}}_{4\mathrm{D}}(k_x,k_y,k_z,k_w)\otimes \H^{\mathrm{D}}_{2\mathrm{D}}(k_u,k_v),\label{H-D-6D}
\end{equation}
where
\begin{equation}
	\begin{split}
	\H^{\mathrm{D}}_{2\mathrm{D}}(k_u,k_v)=\sin k_u \sigma_1&+\sin k_v\sigma_2\\ +&(m-\cos k_u-\cos k_v)\sigma_3.
	\end{split}
\end{equation}
Here, we take $m=-1$, such that $\H^{\mathrm{D}}_{2\mathrm{D}}$ has the first Chern number $1$. Then, $\H_{6\mathrm{D}}^{\mathrm{D}}(\k)$ has the third Chern number $16$. On the boundary normal to the $u$th direction, a $4$D nodal hyperplane with the second Chern number $16$ is observed in the five-dimensional (5D) boundary BZ~[see Fig.~\ref{fig2}(a)]. In Fig.~\ref{fig2}(b), the nodal hyperplane is broken into four $2$D nodal tori under appropriate perturbations. Two of them are explicitly shown in Fig.~\ref{fig2}(c). Each nodal torus can further burst into four nodal points, each having the unit second Chern number [Fig.~\ref{fig2}(d)].  This is consistent with the total charge of 16 of the original nodal hyperplanes.
{Note that the appropriate perturbations can be found by numerical and analytic methods.}


\section{Discussion}

In this work, we have developed a systematic method to construct higher-dimensional ChIs with large Chern numbers. 
Starting from any two ChIs  $\langle \H_A^{\kappa_A}, C_A^p\rangle $ and $\langle \H_B^{\kappa_B}, C^q_B\rangle$, we can construct their 
tensor-product ChI $\langle \H_{AB}^{\kappa_{A}\star\kappa_B}, C_{AB}^{p+q}\rangle$. The AZ symmetry class of the ChI can be designed according to 
Table~\ref{symmetry-class}. And because of the factor $2$ in $C_{AB}^{p+q}=-2C_A^pC_B^q$, the tensor product naturally gives rise to large higher-order Chern numbers.

It is interesting to compare our tensor product approach with the previous proposals to construct higher-dimensional ChIs in Refs.~\cite{Lohse2018,Zilberberg2018}. In those works, essentially, $\H_{AB}(\k^A,\k^B)=\H_A(\k^A)\otimes I_B+I_A\otimes \H_B(\k^B)$, and the energy spectrum is given by $E_{AB}(\k^A,\k^B)=E_A(\k^A)+E_B(\k^B)$. Hence, it is not clear whether the so-constructed $\H_{AB}(\k^A,\k^B)$ is gapped or not. In comparison, our approach has the advantage that the energy gap is naturally preserved from Eq.~\eqref{Sch-Eq}. With the preserved energy gap, the Chern number of the tensor-product Hamiltonian can be uniquely determined and guaranteed to be nontrivial, which is not possible in previous proposals.

Finally, we comment that the tensor structure of \eqref{tensoring} resembles the Hilbert space of two spins. If each spin is varied in an independent parameter space according to a ChI Hamiltonian, then the two-spin system has a higher-order Chern number in the product parameter space~\cite{ZhaoYu2017njp,ZhaoYu2019prl,TanXS2019prl,TanXS2021prl}. Note that in the variation process, the two spins are not entangled, so it can be readily realized. Nevertheless, in such a setup, all dimensions are realized by internal parameters, and therefore it would be difficult to detect the boundary modes.

{As aforementioned, the experimental setup can be considered in artificial systems, such as photonic/phononic crystals, ultracold atoms, electric-circuit arrays, and mechanical systems. Particularly, in electric-circuit networks, the previous works in Refs.~\cite{Ezawa2019prb,Yu_Zhao_NSR,WangYou_2020,ZhangWX2020prb} have realized higher-dimensional topological systems. Hence, in principle our theory can be experimentally demonstrated by artificial crystals.}
		
\section{ACKNOWLEDGMENTS}

This work is supported by the National Natural Science Foundation of China (Grants No. 11874201 and No. 12174181) and the Singapore Ministry of Education AcRF Tier 2 (MOE2019-T2-1-001).

\appendix

\section{The detailed derivation of the main result in the main text}\label{AppexdixA}

We provide the more detailed derivation for the Chern number of the tensored Hamiltonian in the following. For the integral of Eq.~\eqref{Chern-number} in the main text, the commutation relation $ [F_A,F_B]=0 $ leads us to
\begin{equation}\label{FA-FB}
\begin{split}
(F_A+F_B)^n&=\sum_{j=0}^{n}  \begin{pmatrix}
n\\
j
\end{pmatrix} (F_A)^j(F_B)^{n-j}\\
&=\begin{pmatrix}
n\\
p
\end{pmatrix}(F_A)^p(F_B)^q,
\end{split}
\end{equation}
where $ \begin{pmatrix}
n\\
j
\end{pmatrix}=\frac{n!}{j!(n-j)!} $ is the binomial coefficient. The last 
equality holds since $ (F_A)^j=0 $ as $ j>p $, and $ (F_B)^{n-j}=0 $ as $ n-j>q $, i.e., $ j<p $. 
{We can prove these as follows. Since $F_A=F_{\mu\nu}dk_\mu\wedge dk_\nu/2,$ we have
	\begin{equation}
		\begin{split}
			(F_A)^j=&\frac{1}{2^j}F_{\mu_1\nu_1}F_{\mu_2\nu_2}\cdots F_{\mu_j\nu_j}\\
			&dk_{\mu_1}\wedge dk_{\nu_1}\wedge \cdots \wedge dk_{\mu_j}\wedge dk_{\nu_j}.
		\end{split}
	\end{equation}
	If $j>p$ with $2p$ the dimension of system $A$, at least two indexes are equal for each term in $ (F_A)^j $. Note that the wedge product is antisymmetric for these indexes, namely, $\cdots\wedge dk_{\mu}\wedge\cdots \wedge dk_\mu\wedge\cdots=0$, and then we obtain $(F_A)^j=0.$ Similarly, we can prove $(F_B)^{n-j}=0$ as $j<p$.}
By considering Eq.~\eqref{FA-FB}, we have Eq.~(8) of the main text as
\begin{equation}\label{eq8}
C_{AB}^{p+q}=\frac{1}{p!q!}\left(\frac{i}{2\pi}\right)^{p+q}\int_{\text{BZ}} \tr[(F_A)^p(F_B)^q].
\end{equation}
By Eq.~(6) of the main text, we have
\begin{equation}\label{Berry-curvature-1}
\begin{split}
F_A&=(f^{A,-}\otimes I^{B,+})\oplus(f^{A,+}\otimes I^{B,-}),\\ F_B&=(I^{A,-}\otimes f^{B,+})\oplus(I^{A,+}\otimes f^{B,-}),
\end{split}
\end{equation}
where $ f^{\alpha,\pm}=f^{\alpha,\pm}_{\mu\nu}dk_{\mu}dk_{\nu}/2 $. It then gives rise to
\begin{equation}
\begin{split}
(F_A)^p(F_B)^q&=\{[(f^{A,-})^p\otimes I^{B,+}]\oplus[(f^{A,+})^p\otimes I^{B,-}]\}\\
&\quad ~\{[I^{A,-}\otimes (f^{B,+})^q]\oplus[I^{A,+}\otimes (f^{B,-})^q]\}\\
&=[(f^{A,-})^p\otimes (f^{B,+})^q]\oplus [(f^{A,+})^p\otimes (f^{B,-})^q],
\end{split}
\end{equation}
by which we have
\begin{equation}\label{trace}
\begin{split}
\tr[(F_A)^p(F_B)^q]&=\tr[(f^{A,-})^p\otimes (f^{B,+})^q]\\
&\quad +\tr[(f^{A,+})^p\otimes (f^{B,-})^q]\\
&=\tr[(f^{A,-})^p]\tr[(f^{B,+})^q]\\
&\quad +\tr[(f^{A,+})^p]\tr[(f^{B,-})^q].
\end{split}
\end{equation}
Substituting Eq.~\eqref{trace} into Eq.~\eqref{eq8}, the Chern number is obtained as
\begin{equation}~\label{Chern-number-epsilon}
\begin{split}
C_{AB}^{p+q}=&\left[\frac{1}{p!}\left(\frac{i}{2\pi}\right)^p\int_{\text{BZ}_A}\tr[(f^{A,-})^p]\right]\\
&\left[\frac{1}{q!}\left(\frac{i}{2\pi}\right)^q\int_{\text{BZ}_B}\tr[(f^{B,+})^q]\right]\\
&+\left[\frac{1}{p!}\left(\frac{i}{2\pi}\right)^p\int_{\text{BZ}_A}\tr[(f^{A,+})^p]\right]\\
&\quad \left[\frac{1}{q!}\left(\frac{i}{2\pi}\right)^q\int_{\text{BZ}_B}\tr[(f^{B,-})^q]\right]\\
=&C_{A,-}^pC_{B,+}^q+C_{A,+}^pC_{B,-}^q,
\end{split}
\end{equation}
where we have used the definition of Chern number in the last step. Here, $ C_{\alpha,\pm}^j $ is the $ j $th Chern number for the conduction and valence bands of system $ \alpha $, respectively. 


Consequently, by considering that the Chern number is defined over the valence bands as $ C^j_{\alpha}\equiv C^j_{\alpha,-}=-C^j_{\alpha,+} $, Eq.~\eqref{Chern-number-epsilon} gives rise to the main result of the main text as
\begin{equation}\label{main-result}
C_{AB}^{p+q}=-2C_{A}^pC_{B}^q.
\end{equation}
By this main result, we classify the tensor product of the Hamiltonians in classes, C, D, AI, and AII with nontrivial topology in Table~\ref{Topological-Invariant}. 

\section{The calculations of Chern numbers}\label{AppexdixB}

The $ n $-th Chern number defined over $ 2n $D BZ is given as
\begin{equation}\label{nth-CN}
C^n=\frac{n!}{(2n+1)!(2\pi i)^{n+1}}\int_{\mathbb{R}\times\text{BZ}}\tr(GdG^{-1})^{2n+1},
\end{equation}
where $ G $ is the imaginary Green's function. Then, for $ n=2 $, the second Chern number of $\H_{\mathrm{4D}}^\mathrm{AI}(\k)=\H_{\mathrm{2D}}^\mathrm{C}(k_x,k_y)\otimes \H_{\mathrm{2D}}^\mathrm{C}(k_z,k_w)$ (Eq.~\eqref{H-AI-4D} of the main text) can be directly calculated by 
\begin{equation}
\begin{split}
&C_{AB}^2=-\int_{\mathbb{R}\times T^{2n}} \frac{d\omega d^4k}{480\pi^3 i}\epsilon^{\mu\rho\lambda\sigma\tau}\\
&\mathrm{tr}(G\partial_\mu G^{-1}G\partial_\rho G^{-1}G\partial_\lambda G^{-1}G\partial_\sigma G^{-1}G\partial_\tau G^{-1}),
\end{split}\label{2ndChernnumber}
\end{equation} 
where $G(\omega,k_x,k_y,k_z,k_w)=(i\omega-\H_{\mathrm{4D}}^{\mathrm{AI}})^{-1}$, $ \omega $ is integrated over $ \mathbb{R} $, and $ \epsilon $ is the Levi-Civita tensor. By substituting Eq.~\eqref{H-AI-4D} of the main text with $ m=1 $, the second Chern number is obtained as $ C_{AB}^2=-8 $. For the second example of Eq.~\eqref{H-D-6D} in the main text with $ n=3 $, the third Chern number is readily obtained as $ C^3=16 $ for $ m=-1 $.

\section{Topological charges of nodal points}\label{AppexdixC}

To derive the topological charge of the nodal point, we choose a $ (2n-2) $-sphere $ S^{2n-2} $ in the $ (2n-1) $D surface BZ enclosing it provided that the energy spectrum is gapped for all points of the sphere. The topological charge of the nodal point is just the $ (n-1) $th Chern number over this sphere. For example, for a nodal point in the 3D surface BZ, the topological charge can be obtained by the integration over $ \mathbb{R}\times S^{2}  $ as
\begin{equation}
\nu=-\int_{\mathbb{R}\times S^{2}} \frac{d\omega d^2 k}{24\pi^2}\epsilon^{\mu\rho\lambda}\mathrm{tr}(G\partial_\mu G^{-1}G\partial_\rho G^{-1}G\partial_\lambda G^{-1}),\label{Chernnumber}
\end{equation}
where $G=(i\omega-\H_{\mathrm{edge}}^{\mathrm{3D}})^{-1}$ with $ \H_{\mathrm{edge}}^{\mathrm{3D}} $ the effective Hamiltonian of the 3D edge states.

\section{Tensor product of two Hamiltonians in class D}\label{AppexdixD}

In addition to the two examples in the main text, we consider the tensor product of two Dirac models here. It serves as another good example for explaining the basic methods and physics of the main text. First, we consider a simple 2D Dirac model as
\begin{equation}\label{H-D}
\begin{split}
\H_{\text{2D}}^{\text{D}}(k_x,k_y,m)=&\sin k_x\sigma_1+\sin k_y\sigma_2\\
&+(m-\cos k_x-\cos k_y)\sigma_3,
\end{split}
\end{equation}
which belongs to class D with the particle-hole symmetry represented as
\begin{equation}
\hat{\mathcal{C}}_1=\sigma_1\II\K.
\end{equation}
It is readily obtained that the first Chern number is $ C^1=0,\pm1 $ for gapped phases. Then, there is one chiral edge state for the nontrivial phase. 

To obtain the effective Hamiltonian of the open boundary perpendicular to the $x$ direction, we apply the method of inverse Fourier transformation for $k_x$ to get the first quantized Hamiltonian in real space. Note that this method can be generalized to other cases. Then, the Hamiltonian in Eq.~\eqref{H-D} is transformed as
\begin{equation}
\begin{split}
\H_\mathrm{D}(k_y)=&\sin k_y\sigma_2+\frac{1}{2i}(S_x-S_x^\dagger)\sigma_1
\\
&+\left[m-\cos k_y-\frac{1}{2}(S_x+S_x^\dagger)\right]\sigma_3.
\end{split}
\end{equation}
Here, we have used the substitution $ \sin k_x\rightarrow -i(S_x-S_x^\dagger)/2 $ and $ \cos k_x=(S_x+S_x^\dagger)/2 $, where $S_x$ is the translation operators along $ x$ satisfying
\begin{equation}
S_x|i\rangle=|i+1\rangle,\quad S_x^\dagger|i\rangle=|i-1\rangle
\end{equation}
with integer $i$ the lattice site along $x$. Accordingly, for the infinite system, it can be represented as
\begin{equation}
S_x=\begin{pmatrix}
\ddots&\vdots&\vdots&\vdots&\vdots&\reflectbox{$\ddots$}\\
\ddots&0&0&0&0&\cdots\\
\cdots&1&0&0&0&\cdots\\
\cdots&0&1&0&0&\cdots\\
\cdots&0&0&1&0&\cdots\\
\reflectbox{$\ddots$}&\vdots&\vdots&\vdots&\vdots&\ddots
\end{pmatrix}, S_x^\dagger=\begin{pmatrix}
\ddots&\vdots&\vdots&\vdots&\vdots&\reflectbox{$\ddots$}\\
\ddots&0&1&0&0&\cdots\\
\cdots&0&0&1&0&\cdots\\
\cdots&0&0&0&1&\cdots\\
\cdots&0&0&0&0&\cdots\\
\reflectbox{$\ddots$}&\vdots&\vdots&\vdots&\vdots&\ddots
\end{pmatrix}.
\end{equation}
In the presence of the open boundary, the translation operator satisfies $ \bar{S}_x^\dagger\ket{0}=0 $ and $ \bar{S}_x\ket{i}=\ket{i+1} $, which can be represented as
\begin{equation}
\bar{S}_x=\begin{pmatrix}
0&0&0&0&\cdots\\
1&0&0&0&\cdots\\
0&1&0&0&\cdots\\
0&0&1&0&\cdots\\
\vdots&\vdots&\vdots&\vdots&\ddots
\end{pmatrix}, \bar{S}_x^\dagger=\begin{pmatrix}
0&1&0&0&\cdots\\
0&0&1&0&\cdots\\
0&0&0&1&\cdots\\
0&0&0&0&\cdots\\
\vdots&\vdots&\vdots&\vdots&\ddots
\end{pmatrix}
\end{equation}
for the half-infinite system with open boundary at $ i=0 $. Hence, the Hamiltonian for the open boundary condition is now written as
\begin{equation}
\begin{split}
\widehat{\H}_0(k_y)=&\sin k_y\sigma_2+\frac{1}{2i}(\bar{S}_x-\bar{S}_x^\dagger)\sigma_1\\
&+\left[m-\cos k_y-\frac{1}{2}(\bar{S}_x+\bar{S}_x^\dagger)\right]\sigma_3.\label{half-infinte-H}
\end{split}
\end{equation}
We now solve the Schr\"{o}dinger equation for the Hamiltonian in Eq. (\ref{half-infinte-H}) by considering the ansatz of the wave function as 
\begin{equation}~\label{ansatz}
\ket{\psi_{\k}}=\sum_{i=0}^\infty\lambda^i\ket{i}\otimes\ket{\xi_{\k}},
\end{equation}
with $|\lambda|<1$ for the states localized at the boundary. Inside the bulk with $\ket{i\geq 1}$, the Schr\"{o}dinger equation reads
\begin{equation}
\begin{split}
\bigg[&\sin k_y\sigma_2+\frac{1}{2i}(\lambda^{-1}-\lambda)\sigma_1\\
&+\left(m-\cos k_y-\frac{1}{2}(\lambda^{-1}+\lambda)\right)\sigma_3\bigg]\ket{\xi}=\mathcal{E}\ket{\xi},\label{i>0}
\end{split}
\end{equation}
while at $|i=0\rangle$, it reads
\begin{equation}
\bigg[\sin k_y\sigma_2-\frac{1}{2i}\lambda\sigma_1
+\left(m-\cos k_y-\frac{1}{2}\lambda\right)\sigma_3\bigg]\ket{\xi}=\mathcal{E}\ket{\xi}.\label{i=0}
\end{equation}
The difference of Eqs.~\eqref{i>0} and \eqref{i=0} is
\begin{equation}
i\sigma_1\sigma_3\ket{\xi}=\ket{\xi}.\label{projector-kernel}
\end{equation}
This equation implies that the edge state is just the eigenstate of $ i\sigma_1\sigma_3 $ with eigenvalue $ 1 $. Thus, to obtain the effective Hamiltonian of the edge state, we can project the Hamiltonian by the projector
\begin{equation}
\Pi=\frac{1}{2}(1+i\sigma_1\sigma_3)=\frac{1}{2}(1+\sigma_2).
\end{equation}
Applying the projector to Eq.~\eqref{i=0}, we have
\begin{equation}
\sin k_y\sigma_2\ket{\xi}=\mathcal{E}\ket{\xi}.\label{projector-i=0}
\end{equation}
By Eq.~\eqref{projector-kernel}, the difference of Eqs.~\eqref{i=0} and \eqref{projector-i=0} produces
\begin{equation}
\lambda=m-\cos k_y.
\end{equation}
For the validity of the ansatz in Eq.~\eqref{ansatz}, $ k_y $ is constrained by $ |\lambda|<1 $, i.e.,
\begin{equation}\label{restriction}
|m-\cos k_y|<1.
\end{equation}
Then, the effective Hamiltonian for boundary states is obtained as
\begin{equation}
\begin{split}
\H_{\mathrm{edge}}^{\mathrm{1D}}(k_y)&=\Pi\H_D(k_x,k_y)\Pi=\sin k_y\sigma_2\Pi\\
&=\sin k_y\Pi\Rightarrow \H_{\mathrm{edge}}^{\mathrm{1D}}(k_y)=\sin k_y.
\end{split}\label{effective}
\end{equation}
Similarly, the effective Hamiltonian for another boundary is
\begin{equation}
\H_{\mathrm{edge}}^{\mathrm{1D'}}=-\sin k_y.
\end{equation} 
The Fermi point of the edge state locates at $ k_y=0 $.

\begin{figure}[t]
	\includegraphics[scale=0.38]{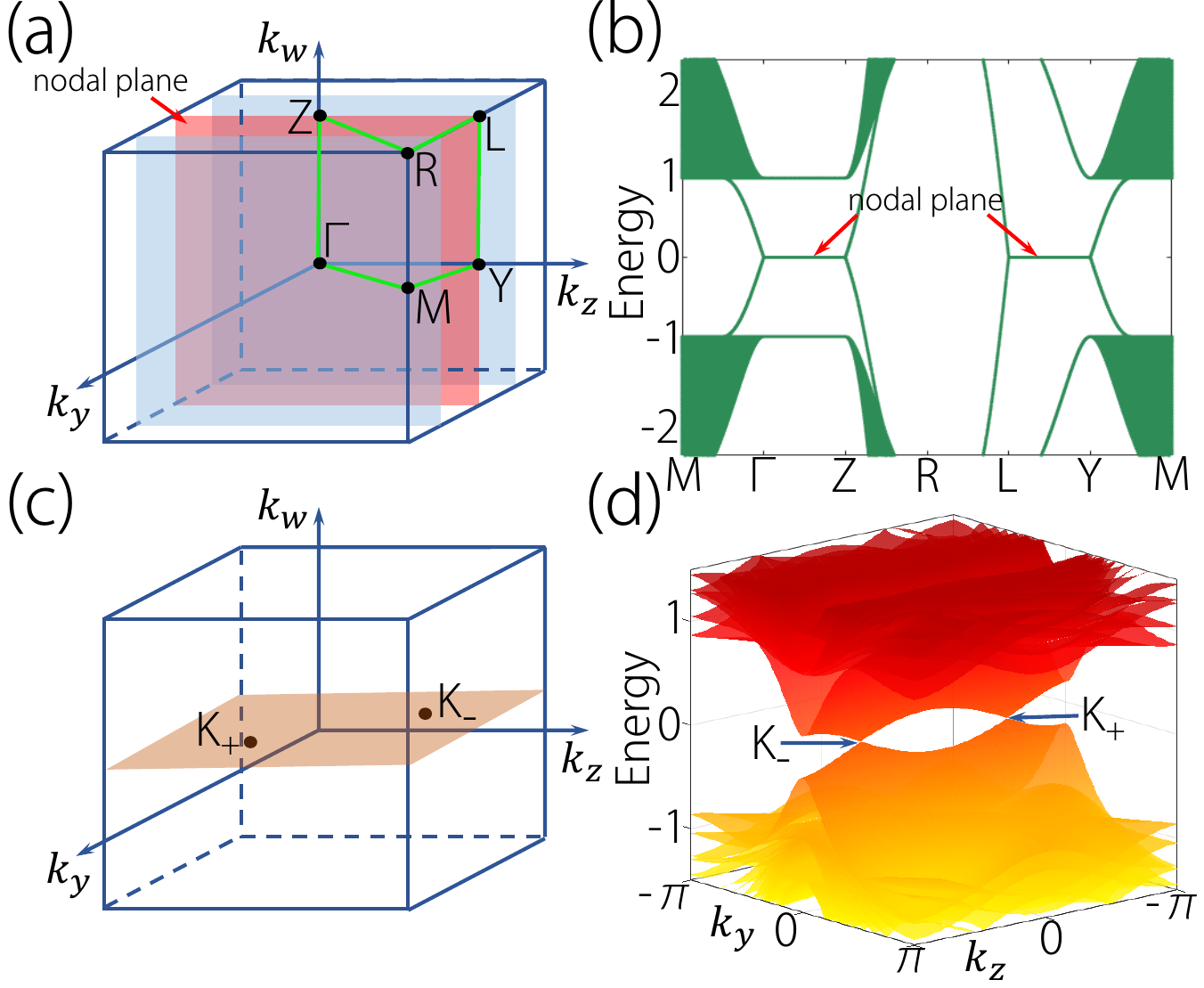}
	\caption{ (a) The 3D surface BZ for the Hamiltonian in Eq.~\eqref{H-AI} with the open boundary normal to the $ x $ direction. The nodal plane (red) is enclosed by two planes (blue). (b) The energy spectrum over the green line of (a). (c) The nodal plane is dissolved into two nodal points $ K_{\pm} $ by the perturbation $ \Delta\H$. (d) The energy spectrum of the boundary states over the plane cutting through two nodal points as shown in (c). Here, we take $ m_1=m_2=1 $ and $ \alpha=0.2 $ in the calculation.}~\label{fig3}
\end{figure}

By tensoring two copies of the Hamiltonian in Eq.~\eqref{H-D} together, we obtain
\begin{equation}\label{H-AI}
\H^{\text{AI}}_{\text{4D}}(\k)=\H^{\text{D}}_{\text{2D}}(k_x,k_y,m_1)\otimes\H^{\text{D}}_{\text{2D}}(k_z,k_w,m_2),
\end{equation} 
which belongs to class AI according to Tab.~\ref{Topological-Invariant}. Time reversal symmetry is now represented as
\begin{equation}
\TT=(\sigma_1\otimes\sigma_1)\II\K
\end{equation}
with $ \TT^2=1 $. When $ m_{1,2}\in(-2,0)\cup(0,2) $, by using Eq.~\eqref{2ndChernnumber}, we obtain the second Chern number as $ C^2_{AB}=-2 $ for $ m_1m_2>0 $ and $ C^2_{AB}=2 $ for $ m_1m_2<0 $ in agreement with the main result in Eq.~\eqref{main-result}.

To investigate the boundary state, we take $ m_1=m_2=1 $ for which $ C_2[\H^{\text{AI}}_{\text{4D}}]=-2 $. By considering the open boundary perpendicular to $ x $, the effective Hamiltonian of its 3D surface state can be directly obtained as
\begin{equation}
\H_{\text{surf}}(k_y,k_z,k_w)=\sin k_y\H^{\text{D}}_{\text{2D}}(k_z,k_w),
\end{equation}
from Eq.~\eqref{effective}. From the constraint of Eq.~\eqref{restriction}, $ k_y\in(-\pi/2,\pi/2) $. Obviously, it hosts a nodal plane $ k_y=0 $ as denoted by the red plane of Fig.~\ref{fig3}(a). The first Chern numbers of the two blue planes in Fig.~\ref{fig3}(a) enclosing the nodal plane are $ \pm1 $, from which we obtain the topological charge of the nodal plane as $ -2 $. It is consistent with the second Chern number. Fig.~\ref{fig3}(b) shows the energy spectrum over the green line in Fig.~\ref{fig3}(a). However, this nodal plane is not topologically stable, which can be dissolved into nodal points by symmetry-allowed perturbation without closing the energy gap. To do this, we introduce a time-reversal-invariant term to $ \H^{\text{AI}}_{\text{4D}}(\k) $ in Eq.~\eqref{H-AI} as 
\begin{equation}
\Delta \H=\alpha\sigma_2\otimes\sigma_1.
\end{equation}
Obviously, $ [\TT,\Delta \H_1]=0 $. If the parameter $ \alpha $ is small enough, the energy gap cannot be closed which guarantees $ C^2_{AB}=-2 $. Using the same method above, the effective Hamiltonian of the surface state is obtained as
\begin{equation}
\H_{\text{surf}}(k_y,k_z,k_w)=\sin k_y\H^{\text{D}}_{\text{2D}}(k_z,k_w)+\alpha\sigma_1,
\end{equation}
which has two nodal points at
\begin{equation}
K_{\pm}=(\pm \arcsin \alpha,\mp\pi/2,0),
\end{equation}
as shown in Fig.~\ref{fig3}(c). The topological charges of both nodal points are obtained as $ -1 $ from Eq.~\eqref{Chernnumber}, which is consistent with the topological charge of the nodal plane without the perturbation. Fig.~\ref{fig3}(d) shows the energy spectrum over the brown plane cutting through the two nodal points in Fig.~\ref{fig3}(c).

{\section{Chiral Symmetry Constrains the Chern Number to be Zero}\label{AppendixE}
\textbf{}
If a system has a chiral symmetry $\Gamma$, the Hamiltonian $\H(\k)$ anticommutes with $\Gamma$, \emph{i.e.}, $\{\H(\k),\hat{\Gamma}\}=0$. Since $\hat{\Gamma}^2=1$ and $\hat{\Gamma}=\hat{\Gamma}^\dagger$, the Hamitonian can be always transformed into block anti-diagonal form, namely
\begin{equation}
	\H(\k)=\begin{pmatrix}
		0&q^\dagger(\k)\\
		q(\k)&0
	\end{pmatrix},
\end{equation}
where $q(\k)$ is an $N\times N$ invertible matrix for each $\k$. Hence, we can always find globally well-defined valence states over the whole BZ as
\begin{equation}
	\ket{\k,n}=\frac{1}{\sqrt{2}}\begin{pmatrix}
		-v_n\\
		q(\k)v_n
	\end{pmatrix},
\end{equation}
where $n=1,...,N$, and $v_n$ is a $N$-vector with all entries being zero expect the $n$th being 1. It is known that the Chern number is an obstruction to find globally well-defined wavefunctions over the whole BZ, thus the chiral symmetry must constrain the Chern number to be zero. Alternatively, one can directly calculate the Berry connection as $A_\mu(\k)=\frac{1}{2}q^\dagger(\k) \partial_\mu q(\k)$ from the valence states above. Then, the Berry curvature is given by $F_{\mu\nu}=\frac{1}{4}(\partial_\mu q^\dagger \partial_\nu q-\partial_\nu q^\dagger \partial_\mu q)$. As a result, $F=\frac{1}{2}F_{\mu\nu}dx^\mu\wedge dx^\nu=\frac{1}{4}dq^\dagger dq$. Hence, $(F)^n=\frac{1}{4^n}d[q^\dagger dq (dq^\dagger dq)^{n-1}]$ is a total derivative. Consequently, the $n$th Chern number over the Brillouin torus vanishes by Stokes' theorem. }

\bibliographystyle{apsrev}
\bibliography{chern-number}

\end{document}